\documentclass[prl,aps,psfig,preprintnumbers,amsmath,amssymb,showpacs]{revtex4}
\usepackage{graphicx}\usepackage{epsfig}
\usepackage{color}

\begin{document}
\title{Thermodynamics of clusterized matter}

\author{Ad. R. Raduta$^{1}$}
\author{F. Gulminelli$^{2}$ }

\affiliation{$^{1}$~NIPNE, Bucharest-Magurele, POB-MG6, Romania,\\
$^{2}$~LPC (IN2P3-CNRS/Ensicaen et Universit\'e), 
F-14076 Caen c\'edex, France}

\begin{abstract}
Thermodynamics of clusterized matter is studied in the framework of
statistical models with non-interacting cluster degrees of freedom.
At variance with the analytical Fisher model, exact Metropolis simulation results
indicate that the transition from homogeneous to clusterized matter lies along
the $\rho=\rho_0$ axis at all temperatures and the
limiting point of the phase diagram is not a critical point even if the surface energy vanishes 
at this point.
Sensitivity of the inferred phase diagram to the employed statistical framework in
the case of finite systems is discussed by considering the grand-canonical and 
constant-pressure canonical ensembles.  
A Wigner-Seitz formalism in which the fragment charge is
neutralized by an uniform electron distribution allows to build
the phase diagram of neutron star matter as predicted by the cluster model.

\end{abstract}

\today

\maketitle

\section{I. Introduction}

Droplet models constitute a suitable tool to study structural ordering 
transitions. Freezing, condensation and glassiness together with 
metallic alloys phase transitions illustrate both the phenomena diversity
and the multitude of domains which benefit from them \cite{binder_clus}.

At sub-atomic scale, clusterized nuclear matter is expected to be present 
in the cores of exploding supernovae \cite{ohnishi,watanabe,mishustin,napolitani}
and in the inner crust of neutron stars 
\cite{ravenhall,lattimer,lattimer_swesty,watanabe2,horowitz,hartreefock,maruyama,constanca}
where the interplay between surface and Coulomb energy is responsible for the occurrence
of exotic pasta phases at densities close to the saturation density of nuclear matter.

{
It is conjectured that clusterized nuclear matter is formed also at the
break-up stage of multifragmentation reactions, i.e. the simultaneous
multi-particle decay of excited matter produced in energetic heavy ion reactions.
It is this scenario 
that persuaded Finn {\it et al.} \cite{finn,zamick} to address
nuclear  multifragmentation is terms of a critical evaporation according to the
Fisher cluster theory \cite{fisher-physics} as early as the '80s.
The approach was daring the more as a parallelism among a neutral infinite system
and an electrically-charged small one is far from being trivial. 
Nonetheless, more and more refined multifragmentation 
data continued to be scrutinized for
criticality signals over more than two decades
\cite{campi,gilkes,elliott,dagostino,texas,leneindre}.
Most of the results tend to confirm the criticality scenario  
and point-out that a first-order phase transition occurs as well
\cite{dagostino2002,indra_cv,indra_corr,indra_tracking,indra_Zmax,multics_Zmax}.
However a real proof of criticality when analyzing 
a finite system like a fragmenting nucleus would need finite size scaling techniques,
and these techniques are not experimentally accessible as the size of the system cannot be varied freely 
in nuclear fragmentation.

From a theoretical point of view, 
several statistical models with cluster degrees of freedom
\cite{smm,mmmc,randrup,mmm} have been proposed 
and proved able to describe a wealth
of experimental data in great detail.
This opens the possibility of addressing the issue of criticality
from a theoretical point of view. However past studies have only 
concerned finite systems \cite{prl2001,coulomb,crit}, while the thermodynamic limit is needed to observe 
non-ambiguous signals of criticality. In this respect
these results are not conclusive.
 

The aim of the present paper is to investigate the phase diagram 
and the critical behavior of an infinite neutral clusterized system.
To reach this goal we undertake the cluster model but, at variance with 
its standard analytical representation commonly known as
the Fisher model \cite{fisher-physics}, we solve it 
exactly via a Metropolis simulation.
This approach will allow us to see to what extent the analytical
approximation stands and, secondly, to investigate how the different physical ingredients act on the thermodynamical properties. 
More precisely, we shall focus 
on effects of free volume correction, fragment compressibility, internal
excitation, translational degrees of freedom and finite size effects.
We shall additionally show that the cluster model can present first and
second order phase transitions but the limiting point of the phase diagram is never a critical point even if 
the surface energy is made vanish at this point.
This implies that either the cluster model in the different versions presented in the literature is not appropriate 
to describe nuclear fragmentation, or that this latter
phenomenon cannot be interpreted as critical.

}

The paper is organized as follows.
Sec. II discusses the thermal and phase properties of nuclear matter as obtained by an exact
Metropolis Monte-Carlo model with respect to the predictions of the analytical Fisher model. 
Sec. III analyzes the sensitivity of small systems to the considered thermodynamical framework, 
while the phase diagram of neutron star matter in the Wigner-Seitz
approximation is presented in Sec. IV. Conclusions are drawn in Sec. V.

\section{II. The cluster model}

It is theoretically well settled that uniform nuclear matter is instable 
to density fluctuations of finite wavelength in a wide region of densities and temperatures,
leading to spontaneous cluster formation \cite{chomaz,rmf_inst}. 
The same is true for stellar matter \cite{pethick,camille,vidana}, 
and the physical properties at the crust-core transition point 
are of special interest for astrophysical purposes \cite{baoanli}.
A description of the thermodynamical properties of these systems can be obtained introducing 
a statistical model with cluster degrees of freedom.
Such an approach was proposed by Fisher
in the late sixties to describe the process of condensation close to a
critical point \cite{fisher-physics}. 

Under the equilibrium assumption,
the statistical approach of a system broken into $N$
pieces reduces the physical problem to the estimation of the number of
microscopic states compatible with the thermodynamical macroscopic constraints.
For instance, 
in the simple case of a 1-component fluid,
if one ignores the internal excitation, position, momenta, spins and parities of the 
clusters,  a configuration will be exclusively defined by fragment sizes,
\begin{equation}
k:\{A_1,...,A_{N_k}\}.
\end{equation}

Including isotopic, translational and internal degrees of freedom, as it is 
usually done to treat the particular case of nuclear multifragmentation, 
leads to:
\begin{equation}
k:\{A_1, Z_1, e_1,{\vec{r}_1}, {\vec{p}_1}, ...,A_{N_k}, Z_{N_k}, e_{N_k},{\vec{ r}_{N_k}}, {\vec{p}_{N_k}} \},
\label{config}
\end{equation}
where $A_i$, $Z_i$, $e_i$, $\vec{r}_i$  and $\vec{p}_i$ stand for the mass, charge, internal energy, 
center-of-mass position and momentum of each fragment.  

Assuming that it is possible to write down a tractable expression for the
statistical weight of the configuration $W_k$
in the statistical ensemble suitable for the phenomenon under study,
one can express 
the characteristic partition sum $Z$ and any ensemble-averaged observable 
$<X>$ as a function of
the statistical weight of the configuration $W_k$,
\begin{equation}
{\cal Z}=\sum_{(k)} W_k,
\label{eq:Z}
\end{equation}
and, respectively, 
\begin{equation}
<X>=\frac {\sum_{(k)} X^{(k)} W_k}{\sum_{(k)} W_k} .
\label{eq:average_value}
\end{equation}

Many improvements and sophistications can and have 
to be added to describe the nuclear multifragmentation problem \cite{mmm} 
or the stellar equation of state \cite{lattimer_swesty,mishustin}, but 
in this section we shall only focus on one-component clusterized systems 
in the grandcanonical ensemble. This will allow for a direct comparison
between the exact solution of the cluster model within Monte-Carlo methods
and an approximation allowing analytical expressions 
at the thermodynamic limit.   
    
\subsection{A. The analytical Fisher model} 
 
In the grand-canonical ensemble, the generic expression of the partition
function of an infinite one-component system composed of independent clusters reads,
\begin{equation}
{\cal Z_{\beta, \beta\mu}}=\sum_{(k)} \exp{ \left[-\beta \left(E_{tot}^{(k)}-
\mu A_{tot}^{(k)} \right) \right]},
\label{zcluster}
\end{equation}
with,
\begin{eqnarray}
E_{tot}^{(k)}=\sum_{A=1}^{\infty} n_A^{(k)} e(A);\\
A_{tot}^{(k)}=\sum_{A=1}^{\infty} n_A^{(k)} A,
\end{eqnarray}
where $n_A^{(k)}$ is the multiplicity of fragments of size $A$ in the $(k)$-class of events, 
$e(A)$ is the cluster energy functional, and for the moment we ignore the degeneracy 
of the different energy states for each fragment size.
The interest of the Fisher cluster model \cite{fisher-physics} is that the
partition sum can be expressed as a product 
of individual cluster partition sums, 
\begin{equation}
{\cal Z_{\beta, \beta\mu}}=\prod_{A=1}^{\infty} z_{\beta, \beta\mu}(A),
\label{fisher_start}
\end{equation}
where
\begin{eqnarray}
z_{\beta, \beta\mu}(A)&=&\sum_{n_A} \exp \left[  -\beta n_A (e(A)-\mu A)\right] \nonumber \\
&=& 1\pm \exp \left[  -\beta (e(A)-\mu A)\right],
\end{eqnarray}
and the $+$ ($-$) accounts for Fermi (Bose) statistics of the clusters states.  If we consider
the classical Boltzmann limit of a huge number of available states, $n_A\ll 1$ we get,
\begin{equation}
\omega_{\beta, \beta\mu}(A)=\ln z_{\beta, \beta\mu}(A)\approx  \exp \left[  -\beta (e(A)-\mu A)\right] ,
\label{petitz}
\end{equation}
leading to,
\begin{equation}
{\cal Z_{\beta, \beta\mu}}=\prod_{A=1}^{\infty} \sum_{n_A=0}^{\infty} \frac{\omega^{n_A}_{\beta, \beta\mu}(A)}{n_A!} ,
\label{zfisher}
\end{equation}
or, equivalently,
\begin{equation}
\ln {\cal Z_{\beta, \beta\mu}}= \sum_{A=1}^{\infty}\omega_{\beta, \beta\mu}(A) ,
\label{fisher1}
\end{equation}
with $\omega_{\beta, \beta\mu}(A)$ given by Eq. (\ref{petitz}).

Taking now into account that the clusters may be excited and have translational degrees
of freedom $e_n(A,\vec{p})=e_n(A)+p^2/2Am$, Eq. (\ref{fisher1}) becomes,
\begin{eqnarray}
\ln {\cal Z_{\beta, \beta\mu}}&\approx& \sum_{A=1}^{\infty}\int\int
\frac{d^3rd^3p}{h^3}\sum_n g_n(A) 
\exp \left[  -\beta (e_n(A,\vec{p})-\mu A)\right] \label{fisher_approx} \\
&\approx& V \left( \frac{m T}{2 \pi \hbar^2} \right)^{3/2}
\sum_{A=1}^{\infty} A^{3/2} \exp \left[  -\beta (f_{\beta}(A)-\mu A)\right] ,
\label{eq:Zgc_analytic}
\end{eqnarray}
where $g_n(A)$ accounts for the degeneracy of internal state $(n)$, the internal entropy is given by 
$\exp s(A,e)=\sum_n g_n(A) \delta(e-e_n(A))$, and we have made a saddle point approximation, 
\begin{equation}
\int_{-\infty}^{\infty} de \exp \left[ s(A,e) - \beta e \right] \approx f_{\beta} (A) ,
\label{saddle}
\end{equation}
where the cluster free energy at temperature $T=\beta^{-1}$ is given  by
$f_{\beta}=\langle e(A) \rangle_{\beta} -T \langle s(A) \rangle_{\beta}$.

It is important to stress that the original Fisher model (Ref. \cite{fisher-physics}) employs the 
saddle point approximation Eq. (\ref{saddle}), but it does not include translational and internal 
degrees of freedom. For consistency these latter should be accounted directly in the partition sum
Eq. (\ref{zcluster}), see Eq. (\ref{config}). 
If this is done the key expression Eq. (\ref{fisher_start})
is not valid any more, and this is why we include 
these extra degrees of freedom only at the level 
of the single-cluster partition sum $\omega_{\beta, \beta\mu}(A)$, leading to the approximation
Eq. (\ref{fisher_approx}).
 
For nuclear (fermionic) clusters both the average cluster energy and entropy 
can be evaluated in the low temperature Fermi gas approximation
\begin{eqnarray}
\langle e(A) \rangle_{\beta}&=&-B(A)+a_0 A T^2; \label{estar_fisher} \\
\langle s(A) \rangle_{\beta}&=& \left ( 2a_0 T +a_s A^{-1/3} f(T) \right ) A
- \tau \ln A ,
\label{entropy_fisher}
\end{eqnarray}
and to have an analytically solvable model, the fragment binding energy
$B=-e_1$ can be simply parameterized as:
\begin{equation}
B(A)=a_v A -a_s A^{2/3}. 
\label{eq:B}
\end{equation}
The surface term $a_s A^{2/3}f(T)$ in Eq. (\ref{entropy_fisher}) effectively accounts for the 
entropy increase at finite temperature due to surface excitations. In particular if we choose
for $f(T)$ a smooth function such that $f(T)=\beta$ for $\beta \leq \beta_c$, this naturally
produces a vanishing surface free energy at the critical point $\beta_C=T_C^{-1}$.
Concerning the Fisher topologic factor $\tau \ln A$, it insures a fragment fractal dimension 
at the critical point. The physical meaning of this term, as pointed out in
Ref. \cite{lattimer85},
is related to the treatment of the center of mass degree of freedom close to a critical point.
In such a situation, the available states for the center of mass motion 
have to be reduced respect to those of a free particle, and this can be accounted by reducing 
the exponent associated to the translational energy factor $\propto A^{3/2}$
in Eq. (\ref{eq:Zgc_analytic}).
We have taken $\tau_0=\tau-3/2=2.2$ in order to recover the fractal dimension
of the LG universality class. 
A more sophisticated temperature dependent expression for $\tau$ has been
proposed in Ref. \cite{lattimer85}.

The specific form of Eqs. (\ref{entropy_fisher}) and (\ref{estar_fisher}) insures that the model is 
critical at $T=T_C$, $\mu=\mu_C=-a_v-a_0 T_C^2$. 
Indeed, computing the average multiplicity of clusters of size $A$,
\begin{eqnarray}
\langle n_A \rangle_{\beta,\beta\mu}&=&
\frac{\partial \ln \cal Z_{\beta\mu}}{\partial \left ( \beta \mu \right )}
\equiv \omega_{\beta, \beta\mu}(A)= \nonumber \\
&=& V \left( \frac{m T}{2 \pi \hbar^2} \right)^{3/2}
 A^{3/2} \exp \left[  -\beta (f_{\beta}(A)-\mu A)\right]. 
\end{eqnarray}
One can see that the partition sum of the Fisher model is simply given by
the sum of all the possible cluster multiplicities:
\begin{equation}
\ln {\cal Z_{\beta, \beta\mu}}= \sum_{A=1}^{\infty}  \langle n_A \rangle_{\beta,\beta\mu}.
\label{fisher_end}
\end{equation}

At $T=T_C$, $\mu=\mu_C$ the fragment multiplicity obeys a power law,
\begin{equation}
\langle n_A \rangle_{\beta_C,\beta_C\mu_C}= 
V \left( \frac{m T_C}{2 \pi \hbar^2} \right)^{3/2} A^{-\tau_0}  .
\end{equation}

Since the different thermodynamic observables can be expressed as successive
derivatives of the partition sum, 
this means that all thermodynamic quantities will exhibit a power law behavior
at the approach of $T=T_C$, $\mu=\mu_C$.
In particular the critical exponents $\gamma$, $\alpha$ may be calculated numerically out
of the total particle number,
\begin{eqnarray}
\chi &\propto& |T_C-T|^{-\gamma}, \nonumber \\
&=&\frac{\sigma^2_{N;\beta, {\beta\mu}}}{T}=
\frac{1}{T} \frac{\partial^2 \ln {\cal Z_{\beta, \beta\mu}}}{\partial \alpha^2},
\label{eq:sig2N}
\end{eqnarray}
and, respectively, total energy fluctuation,
\begin{eqnarray}
C_V &\propto & |T_C-T|^{-\alpha}, \nonumber \\
&=&\frac{\sigma^2_{E;\beta, {\beta\mu}}}{T^2} =
\frac{1}{T^2} \frac{\partial^2 \ln {\cal Z_{\beta, \beta\mu}}}{\partial \beta^2}.
\label{eq:sig2E}
\end{eqnarray}

It is easy to see that any value for the topologic exponent $2<\tau_0<3$ will
produce a diverging susceptibility  
$\chi$ and a finite critical density,  
\begin{equation}
\rho_c=\frac{1}{\tau_0-2} \left( \frac{m T_C}{2 \pi \hbar^2} \right)^{3/2},
\label{rhocrit}
\end{equation}

The critical pressure is also immediately calculated as,
\begin{equation}
P_C=T_C \frac{\ln {\cal Z_{\beta, \beta\mu}}|_C}{V} 
=T_C \left(\frac{m T_C}{2 \pi \hbar^2}\right)^{3/2} \frac{1}{\tau_0-1}.
\end{equation}

This formulation together with its canonical and microcanonical counterparts 
have been extensively used to study nuclear multifragmentation
and most of the interesting features of its associated thermodynamics
\cite{mekjian,dasgupta,dasgupta2006,bugaev}: phase transition, critical behavior,
finite size, isospin asymmetry and Coulomb effects,
often with contradictory results.

\subsection{B. The Metropolis Monte-Carlo approach} 

A complementary way to solve our problem is by employing a Metropolis Monte-Carlo
investigation of the configuration space.
While having the obvious disadvantage of being analytically non-tractable,
this method has the advantage of being solvable without approximations for any finite 
number of particles.
Let us start again from the grancanonical partition sum of the cluster model
for a finite system of volume $V$:

\begin{equation}
{\cal Z_{\beta, \beta\mu}}=\sum_{(k)} \frac{1}{N_k!} \int \frac{d^3 r_1 \dots 
d^3 r_{N_k} d^3p_1\dots d^3p_{N_k}}{h^{3N_k}}
\exp{ \left[-\beta \left(E_{tot}^{(k)}-\mu A_{tot}^{(k)} \right) \right]},
\label{mmm_start}
\end{equation}
with
\begin{eqnarray}
E_{tot}^{(k)}=\sum_{i=1}^{N_k} e(A_i,\vec{p}_i,n_i);\\
A_{tot}^{(k)}=\sum_{i=1}^{N_k} A_i,
\end{eqnarray}
where again the cluster energy functional consists of kinetic energy and
internal levels: 
$e(A,\vec{p},n)=e_n(A)+p^2/2Am$.
Calculating the phase space integral Eq. (\ref{mmm_start}) simplifies to
\begin{eqnarray}
{\cal Z_{\beta, \beta\mu}}&=&\sum_{(k)} W_k \nonumber \\
&=& \sum_{(k)} \frac{1}{N_k!} V^{N_k} \prod_{i=1}^{N_k}
\left(\frac{m A_i T}{2 \pi \hbar^2}\right)^{3/2} 
\exp \left[  -\beta \left( f_\beta(A_i)- \mu A_i \right) \right],
\label{eq:Wc}
\end{eqnarray}
where we can take the same free energy functional Eqs. (\ref{estar_fisher}, \ref{entropy_fisher}) 
as for the Fisher model in order to compare the two approaches.
In principle inter-fragment interactions should be taken into account in the
form of an excluded volume \cite{mmm}.
This effect is not accounted in the original Fisher model \cite{fisher-physics}
because it is a size-dependent effect
which becomes negligible in the thermodynamic limit of interest here, 
as we will discuss in the next section. 
Therefore we will consider point-like 
particles also in the Monte-Carlo version of the model, Eq. (\ref{eq:Wc}).

In the thermodynamic limit we can write
\begin{equation}
\lim_{V\to \infty} \sum_{(k)} \prod_{i=1}^{N_k} = \prod_{A=1}^{\infty}
\sum_{n_A=0}^{\infty}  .
\end{equation}
This means that, in the absence of the factorials $1/N_k!$ in Eq. (\ref{eq:Wc})
and $1/n_A!$ in Eq. (\ref{zfisher}),
the analytical and Monte-Carlo model would be equivalent at the thermodynamic limit.
Because of the factorials, this equivalence is violated: 
considering the exact counting of all the possible configurations including
their translational and internal degrees of freedom
breaks the independence of the different cluster sizes Eq. (\ref{fisher_start}) 
even if we keep the Fisher hypothesis of non-interacting point-like clusters.
Eq. (\ref{fisher_end}) has then to be considered as an approximation 
of the exact partition sum of the cluster model in the thermodynamic limit. 
The thermodynamic properties of the cluster model
should then be investigated solving numerically Eq. (\ref{eq:Wc}) for a finite size system big enough 
that the independence of the system size is achieved, or that converging quantities can be extracted 
through finite size scaling techniques. 

The spanning of the available configurations in Eq. (\ref{eq:Wc}) is done by proposing successively 
trial configurations 
which are accepted according
to the detailed balance principle:
\begin{equation}
N_k! \, P(k->k') W_k \Delta k=N_{k'}! \, P(k'->k) W_{k'} \Delta k',
\end{equation}
see Ref. \cite{mmm} for details.
Direct access to full thermodynamical information of each statistical 
configuration allows one to trace
phase coexistence from the bimodal structure of the 
different probability density distributions 
\cite{binder,gulminelli-annphys}
and, moreover, to identify to which phase each configuration belongs.

In general and by construction, the double peak structure
of the probability distribution originates from the convex
intruder in the surface of the associated density of states 
as a function of the order parameter of the underlying
first-order phase transition\cite{binder}.
We expect particle and energy density as possible order parameters.
In the grand-canonical case under consideration, 
the probability distribution in the presence of a first order phase transition 
should then be double-peaked with respect
to the extensive observables $A_{tot}$ and $E_{tot}$ conjugated to the
Lagrange multipliers $\beta$ and $\alpha=\beta\mu$.

To allow for a straightforward comparison with the case of nuclear matter, 
in this Section, the following set of parameters will be used:
($a_v$=16 MeV, $a_s$=16 MeV, $a_0$=1/16 MeV$^{-1}$, $T_c$=16 MeV).
To test the results stability to model parameters,
two different expressions have been employed for $f(T)$: \\
(i) $f(T)=T^{-1} \left [ 1 - \left( (T_C^2-T^2)/(T_C^2+T^2)\right ) ^{1.25}\Theta(T_C-T)\right ]$,  and \\ 
(ii) $f(T)=T^{-1} \left [ 1 - \left ( 1 - T/T_C \right )\Theta(T_C-T) \right
]$.
 
\subsection{C. The phase diagram}

Let us consider a system confined into the finite volume $V$ and denote by $A_0$ 
the maximum number of particles corresponding to normal density, $\rho_0=A_0/V$.

\begin{figure}
\begin{center}
\includegraphics[angle=0, width=0.9\columnwidth]{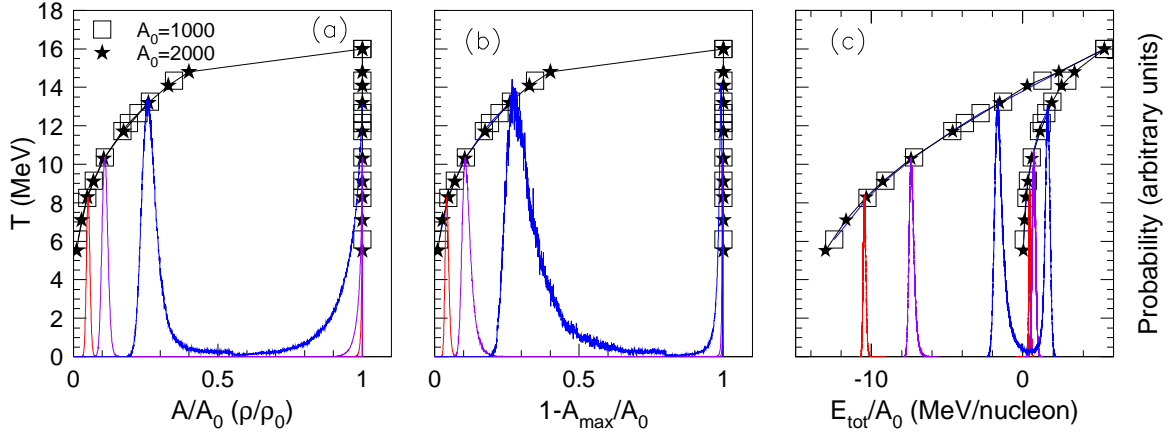}
\end{center}
\caption{(color online) 
Probability distributions of the total number of particles (left),
largest fragment in each event (middle) and
total energy (right) for a one-component cluster model 
with standard Fisher fragment definition
in the grand-canonical ensemble characterized by
$V$=14476.4 fm$^3$ ($A_0$=2000) and
$(T,(\mu/T))$ = (8.3, -2.40), (10.35, -2.20) and (13.19, -2.05), 
where $T$ is expressed in MeV (Metropolis simulation results).
The normalization factors are chosen such that the peak heights
correspond to the coexistence temperature.
The solid star symbols mark the phase diagram in temperature - density (left and
middle panel) and temperature - total energy representations as obtained out
of the bimodality of corresponding distributions.
Open squares mark the phase diagram of a system 
half the size, ($V$=7238.2 fm$^3$ and $A_0$=1000).}
\label{fig:phd_fisher_standard}
\end{figure}

Fig. \ref{fig:phd_fisher_standard} illustrates the probability distributions of 
the total number of particles (left panel) and total energy (right panel) 
for a system characterized by $V$=14476.4 fm$^3$ ($A_0$=2000) at
different values of $(\beta,\alpha)$,
together with the phase diagram (stars) in temperature - total density (left) and
temperature - total energy (right) representations.
To allow superposition of the above plots, the normalization factors of the probability distributions
have been arbitrarily altered such that the peak heights correspond to the 
coexistence temperature represented on the left Y-axis.
To estimate finite size effects, 
the phase diagram of a system two times smaller ($V$=7238.2 fm$^3$
and $A_0$=1000) is represented with open squares.

Given that $A_{tot}$ and $E_{tot}$ are correlated, 
at the thermodynamic limit the phase-coexistence
information provided by the bimodal structure of the these
distributions should be the same. 
Indeed, the phase-coexistence temperature - 
defined by the temperature value for which the peaks belonging to the 
dense and diluted phases have the same height if $\alpha$ is kept constant, 
differs by less than 0.1 MeV 
when one considers $Y(A_{tot})$ or $Y(E_{tot})$ distributions.
This is coherent with a one-dimensional order parameter transition
like ordinary liquid-gas.

It is interesting to observe that correct thermodynamic information is obtained 
also from the probability distributions of the largest cluster $A_{max}$ in each event
(middle panel) as in fragmentation transitions. This is due to the strong correlation of
$A_{max}$ with the total energy deposit
\cite{amax-latticegas,gulminelli-amax,chaudhuri}. 
Even more interesting, plotted as a function of $(1-A_{max}/A_0)$, 
the phase diagram built out of $A_{max}$ bimodality
sits perfectly on the top of the one built analyzing the bimodality of 
$Y(A_{tot})$ distributions. 
Thus, we confirm that $A_{max}$ is a good order parameter in the cluster model. 
This is an encouraging result for the experimental search of the finite nuclei phase diagram
out of the measured $A_{max}$ distributions \cite{indra_Zmax,multics_Zmax}.
 
As one may see in Fig. \ref{fig:phd_fisher_standard}, the shape of the phase
diagram does not depend on the system size. 
The gas-like branch approaches the critical point of the analytical Fisher
model discussed in the previous section.
However, the limiting point characterized by $T_{lim}$=16 MeV and
$\alpha_{lim}$=-2, where by construction the fragment size distribution is an
exact power low, lies on the line $\rho=\rho_0$, which is very different from
the value  $\rho_c \approx 0.07488$ fm$^{-3}$ obtained in the analytical
Fisher model Eq. (\ref{rhocrit}).
The same result has been obtained in Ref. \cite{bugaev}, 
where the phase diagram of a 1-component clusterized system was
calculated analytically out of the Gibbs equations.
This result is not physical and is due to the fact that the cluster model considers all 
nuclear clusters fully incompressible, including 
the asymptotic limit of nuclear matter. 

Within this modelisation, the shape of the phase diagram and in particular its 
limiting temperature are independent of the system size for any $A_0>100$. 
This means that the thermodynamic limit may be addressed directly from finite size systems calculations  
without performing finite size scaling. 
Finite size scaling implies a power law for the size dependence 
of the limiting temperature if this latter corresponds to a critical point, according to 
\begin{equation}
(T_{lim}(L)-T_C) \propto L^{-1/\nu},
\end{equation}
where $L$ is the system linear dimension, $T_C=\lim_{L->\infty} T_{lim}(L)$, and $\nu$
is the critical exponent related to the divergence of the correlation length at the critical point,
$\xi \propto |1-T/T_C|^{-\nu}$. 
The observed independence of  $T_{lim}(L)$ from the system size 
indicates that the correlation length of the infinite system does not diverge
at $T_C$, 
which means that the limiting point is not a critical point.  
This again is at variance with the results of the analytical Fisher model,
and in contradiction with most expectations on nuclear multifragmentation based on statistical
clusters models.

Before leaving this section, we mention that the shape of the phase diagram
proves insensitive to the functional dependence according to which
the surface energy term vanishes while approaching $T_C$.
The solid stars and open squares in Fig. \ref{fig:phd_fisher_standard}
correspond to the case in which $f(T)$ has the form (i).

\subsection{D. Criticality}

The question whether or not this second-order transition point is indeed 
a critical point may be also answered calculating different critical exponents
and checking whether they do verify, together with the input $\tau$, 
the corresponding equality relations \cite{fisher-physics},
\begin{equation}
\frac{\gamma}{(2-\alpha)}=\frac{(3-\tau)}{(\tau-1)}=1-\frac{2\beta}{2-\alpha},
\label{eq:eq_gamma}
\end{equation}
and,
\begin{equation}
\delta=\frac{1}{\tau-2}.
\label{eq:test_delta}
\end{equation}

The full thermodynamic characterization of the system allows the calculation
of $\beta$ and $\delta$ out of the curvature of the coexistence curve,
\begin{equation}
\left(1-\frac{\rho_{coex}}{\rho_C}\right) \propto \left(
1-\frac{T}{T_C}\right)^{\beta},
\label{eq:beta}
\end{equation}
and, respectively, the shape of the critical isotherm,
\begin{equation}
\left( p_C -p\right)|_{T_C} \propto (\rho_C-\rho)^{\delta}|_{T_C},
\end{equation}
which is equivalent through $(p_C-p) \propto \rho_C (\mu_C-\mu)$ to,
\begin{equation}
\left( \mu_C -\mu \right)|_{T_C} \propto (\rho_C-\rho)^{\delta}|_{T_C}.
\label{eq:delta}
\end{equation}

\begin{figure}[b!]
\begin{center}
\includegraphics[angle=0, width=0.98\columnwidth]{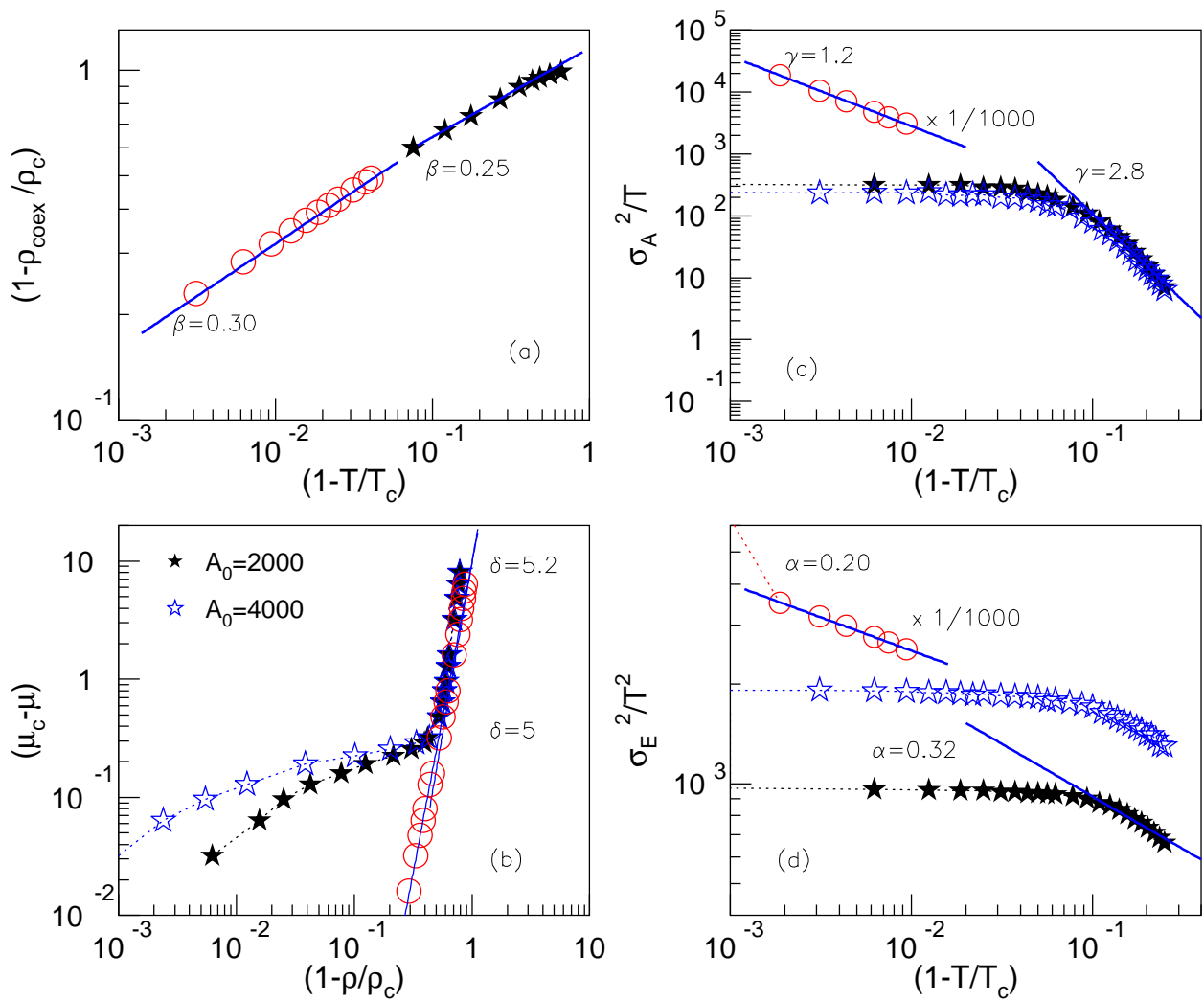}
\end{center}
\caption{(color online) 
Extraction of critical exponents corresponding to a 
one-component clusterized system within a
grand-canonical ensemble out of the thermodynamical properties.
Results corresponding to the analytical model are plotted with open circles,
while stars stand for the results of the Metropolis simulation.
In this last case, 
the maximum number of particles allowed by the fixed volume is $A_0=2000$
(solid stars) and, respectively, $A_0=4000$ (open stars).
Panel a: $\beta$ extracted out of the coexistence curve $\rho-T$;
panel b: $\delta$ extracted out of the critical isotherm;
panel c: $\gamma$ extracted out of total particle number fluctuations;
panel d: $\alpha$ extracted out of total energy fluctuations.
Cases (c) and (d) consider the critical $(\mu/T)$ path. 
}
\label{fig:critexp_thermo_standard}
\end{figure}

The absence of finite size effects in the phase diagram allows direct
extraction of $\beta$ out of the curvature of the coexistence curve, 
Eq. (\ref{eq:beta}). 
The results of the Metropolis simulation for the one-component clusterized system
in the grandcanonical ensemble are plotted in panel (a) of
Fig. \ref{fig:critexp_thermo_standard}
with solid stars along with the predictions of the analytical model, 
represented with open circles. 
Over the explored $(\mu/T)$ domain 
the coexistence curve $\rho_{coex}$  exhibits
the expected power law behavior as a function of $T$ 
in the two models, with slightly different values for the exponents. 

By contrast with this, when approaching the critical point the critical
isotherm deviates from the expected linear behavior and the magnitude of these
deviations increases with the system size, as illustrated in the 
panel (b) of Fig. \ref{fig:critexp_thermo_standard} where full and open
stars correspond to the predictions of the grandcanonical one-component
Metropolis model with $A_0$=2000 and 4000.  This behavior can
be understood taking into account that the calculation of $\rho$ demands the evaluation 
of the average number of particles, a quantity which, like any statistical average,
is naturally affected by the finite size of the system.
The slope of the linear region $\delta$=5.2 is, nevertheless, fairly close
to the predictions of the analytical model such that one may conclude that
the hyperscaling equation (\ref{eq:test_delta}) holds.
 
$\gamma$ and $\alpha$ may be directly extracted out of the fluctuations of
the system total energy and total number of particles via Eqs. 
(\ref{eq:sig2N}) and (\ref{eq:sig2E})
along the critical $\alpha$=-2 path.

The results of the Fisher analytical model shown in panels (c,d) fulfill again perfectly
the hyperscaling relations Eq. (\ref{eq:eq_gamma}), as expected from the
discussion of the previous section.
Since fluctuations should diverge at the approach of a critical point, 
the Monte-Carlo results obtained for a finite system must obviously saturate
at a finite value.
This can be seen in panels c and d, where, irrespective of the system size, 
one may distinguish two regimes for the Monte-Carlo results:
in the temperature domain $0.1 < 1-T/T_C < 0.7$, $\sigma^2_A/T$ and
 $\sigma_E^2/T^2$ are linear with a non-null slope, while for $T$ approaching
$T_C$, the two curves get flat. 
The comparative analyze of $A_0$=2000 and 4000 results shows that neither the
widths of the two intervals, neither the value of the non-null slope depend on
the system size.
The first observation is a first indication that the system might not be critical.
Indeed, if the system were critical, by increasing the system size, the linearity
domain close to the critical point would increase. This is obviously not the
case here.
The second observation is in line with the lack of system size dependence in
the phase diagram and
allows one to extract the critical exponents without
performing finite size scaling. 
The values $\alpha=0.34$ and $\gamma=2.8$ 
obtained over the region where $\ln (1-\rho_{coex}/\rho_C)$ vs. $\ln (1-T/T_C)$ is linear
are so different from the expected values 
$\alpha=0.2$,$\gamma=1.2$ extracted from Eq. (\ref{eq:eq_gamma}) that it is difficult to believe
that this can be a finite size effect.

\begin{table}
\begin{center}
\caption{Critical exponents for a one component clusterized system 
within a grandcanonical ensemble
as obtained from a Metropolis simulation in comparison with results of the
analytical model and the expectations from the Liquid-Gas universality class.}
\begin{tabular}{ccccc}
\hline
\hline
Critical exponent & Calculation method & Monte Carlo  &
Analytical & LG \\
\hline
$\alpha$             & $\sigma_E^2/T^2 \propto |T-T_C|^{-\alpha}$ & 0.34  &
0.2 & 0.1 \\
$\gamma$             & $\sigma_A^2/T \propto |T-T_C|^{-\gamma}$ & 2.8   &  1.2
& 1.2   \\
$\beta$              & $(\rho_C-\rho_{coex}) \propto (T_C-T)^{\beta} $ & 0.25  &
0.33 & 0.33 \\
$\delta$             & $(p_C-p)|_{T_C} \propto (\rho_C-\rho)^{\delta}|_{T_C} $ & 5.2   &
5  & 5 \\
\hline
$\gamma_{fr}$         & $m_2 \propto (T_C-T)^{-\gamma}$ & 2.8   &     & 1.2
 \\
$\beta_{fr}$          & $A_{max} \propto (T_C-T)^{\beta}$ & 0.33  &    & 0.33
 \\ 
$(\beta/\gamma)_{fr}$ & $A_{max}  \propto S_2^{\beta/\gamma}$ & 0.16  &  0.275
& 0.275  \\
\hline
\end{tabular}
\end{center}
\label{table:crit_exp}
\end{table}

The ensemble of these results show that 
criticality is violated in the cluster model even if the surface tension vanishes at the limiting point 
of the first-order phase transition.

In percolation theory over infinite \cite{percolation} 
and small lattices \cite{smalllattice},
critical exponents may be extracted from
$k$-moments of the cluster distributions,
\begin{equation}
m_1=\sum_s s n(s) \propto \left( p-p_C\right)^{\beta},
\label{eq:m1}
\end{equation}
\begin{equation}
m_2=\sum_s s^2 n(s) \propto |p-p_C|^{-\gamma},
\label{eq:m2}
\end{equation}
where $p$, the probability that a lattice site is occupied, is the control
parameter. When dealing with finite systems at $T<T_C$, the largest cluster
assimilated with the liquid phase must be excluded from summations \cite{percolation}.
 
In nuclear multifragmentation, and Eqs. (\ref{eq:m1}) and (\ref{eq:m2})
have been employed in the equivalent forms \cite{campi,gilkes}
\begin{equation}
A_{max} \propto (T_C-T)^{\beta}  
, ~~{\rm for}~~ T<T_C,
\label{eq:amax-T}
\end{equation}
\begin{equation}
m_2 \propto (T_C-T)^{-\gamma}, ~~{\rm for}~~ T<T_C,
\label{eq:m2bis}
\end{equation}
and,
\begin{equation}
A_{max} \propto S_2^{\beta/\gamma},
\label{eq:amax-s2}
\end{equation}
where $ S_2=m_2/m_1$,
and often the distance to the critical point $T-T_C$ has been estimated 
from the fragment multiplicity $M$.
Eq. (\ref{eq:amax-s2}) is particularly easy to handle as
it does not require data sorting according to $T$, a quantity very difficult
to infer experimentally. 

\begin{figure}[b!]
\begin{center}
\includegraphics[angle=0, width=0.6\columnwidth]{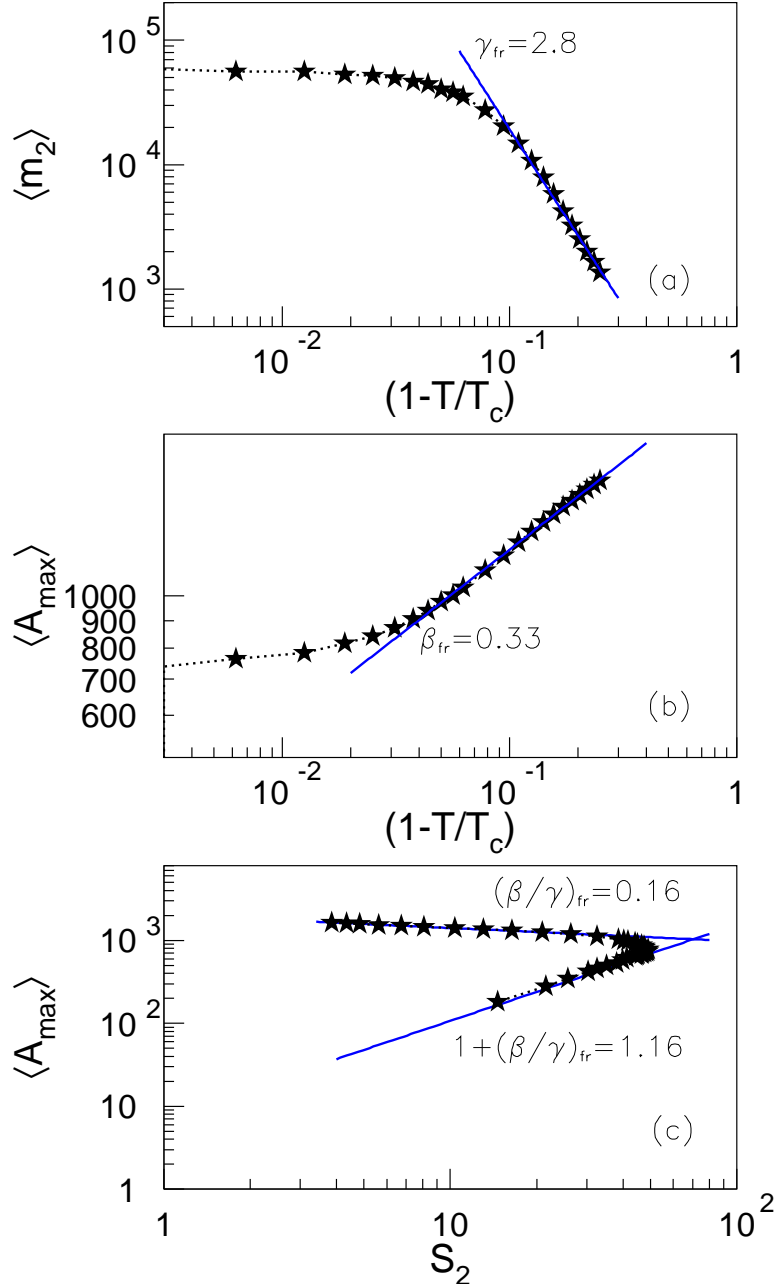}
\end{center}
\caption{(color online) 
Extraction of critical exponents $\gamma$ (Eq. (\ref{eq:m2}), top panel), 
$\beta$ (Eq. (\ref{eq:amax-T}), middle panel)
and $\beta/\gamma$ (Eq. (\ref{eq:amax-s2}), bottom panel) 
out of fragment properties along the critical $(\mu/T)$=-2 path
in the case of a one-component clusterized system within a grand-canonical ensemble.
The maximum number of particle allowed by the fixed volume is $A_0=2000$.
}
\label{fig:critexp_frag_standard}
\end{figure}

Fig. \ref{fig:critexp_frag_standard} investigates to what extent the parallelism
between percolation on lattice and clusterized matter holds.
The evolution of the second moment of the cluster distribution and
the largest fragment in each event as a function of $(1-T/T_C)$ for $T < T_C$
are plotted along the critical $\alpha$=-2 path (panels a and b).
After eliminating the region where divergences are rounded-off by
finite size effects, 
one gets $\gamma_{fr}=2.8$ and $\beta_{fr}$=0.33. 
The result for $\gamma$ is in excellent agreement with the fluctuation
estimation, and again very different from the 
value $\gamma=1.2$ one should get to have thermodynamically consistent exponents. 
The slight disagreement between the $\beta$ values may be surprising
considering that we have observed that $A_{max}$ 
is an excellent order parameter in the cluster model. However it is interesting to note that,
at variance with all other procedures here
employed, Eq. (\ref{eq:amax-T}) proves sensitive to the way in which fragment
surface energy vanishes while approaching the limiting temperature.
For instance for $f(T)=T^{-1} \left[1 - \left(1- T/T_C\right) \Theta(T_C-T) \right] $,
Eq. (\ref{eq:amax-T}) leads to $\beta_{fr}$=0.25.
The predictions of Eq. (\ref{eq:amax-s2}) are illustrated in panel c for
12 MeV $<T<$ 24 MeV along $\alpha$=-2.
Liquid-like and gas-like branches of the Campi plot give $(\beta/\gamma)_{fr}$=0.16.
This number is not compatible with criticality which would require $(\beta/\gamma)=0.275$, but it
is in striking agreement with the values reported in the experimental multifragmentation 
literature \cite{campi,dagostino}. 
This result is surprising since deviations might have been expected in data
due to the Coulomb interaction.

Before leaving this section let us mention that, if we would only dispose of cluster 
observables as in the experimental case of multifragmentation, the results for $(\beta/\gamma)_{fr}$,
$\gamma_{fr}$ and $\beta_{fr}$ from Eqs. (\ref{eq:amax-T}), (\ref{eq:m2bis}) and (\ref{eq:amax-s2}), 
would make us erroneously conclude that the exponents are thermodynamically consistent and 
the system is critical. The computation of thermodynamical quantities as in Figure \ref{fig:critexp_thermo_standard} above is necessary to realize the violation of criticality of the model. Therefore there is no contradiction between the fact that the cluster model is not critical and the fact that this model is so successful in reproducing experimental data which have been qualified as critical.

The values of the above discussed critical exponents are summarized in Table I 
together with the employed equations.

\subsection{E. Temperature dependent bulk energy}

\begin{figure}
\begin{center}
\includegraphics[angle=0, width=0.9\columnwidth]{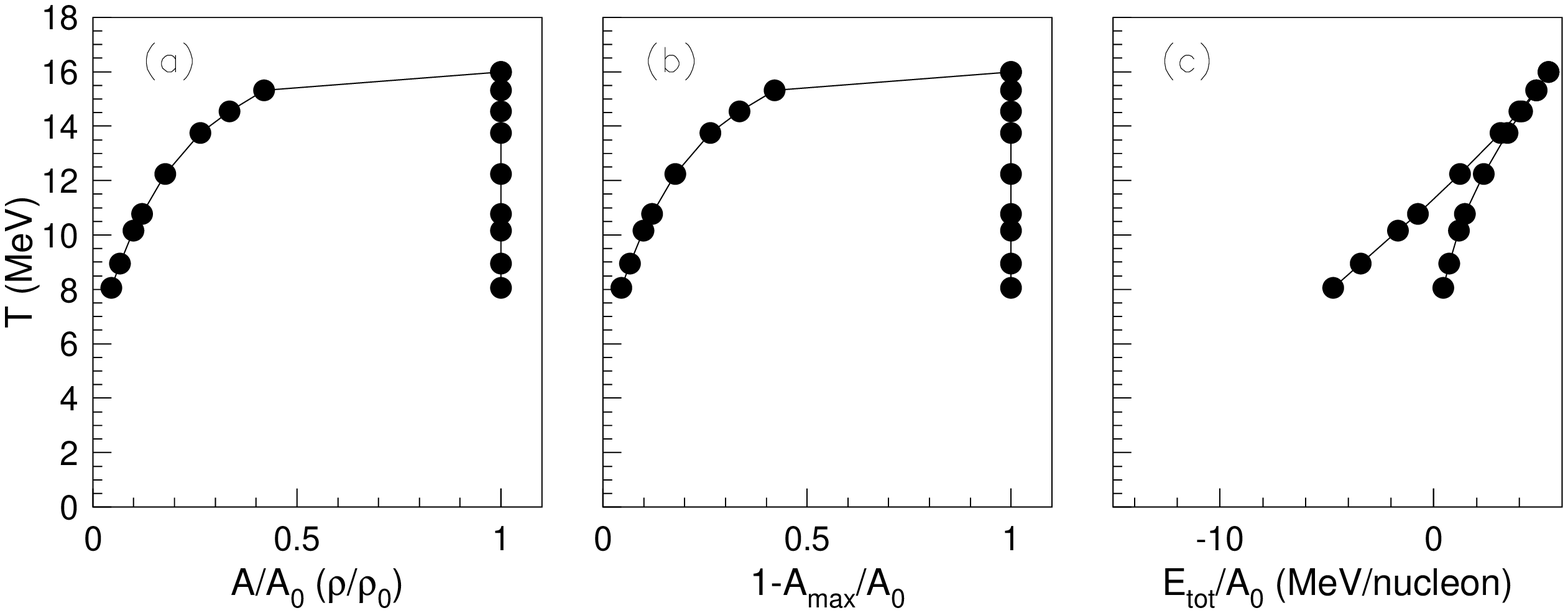}
\end{center}
\caption{
The phase diagram in temperature - density (left and
middle panel) and temperature - total energy representations as obtained out
of the bimodality of corresponding distributions
for a system characterized by $V$=14476.4 fm$^3$ ($A_0$=2000)
within the grand-canonical ensemble.
At variance with standard fragment definition, here we assume a
temperature dependence of the bulk energy.}
\label{fig:phd_fisher_t}
\end{figure}

\begin{figure}
\begin{center}
\includegraphics[angle=0, width=0.98\columnwidth]{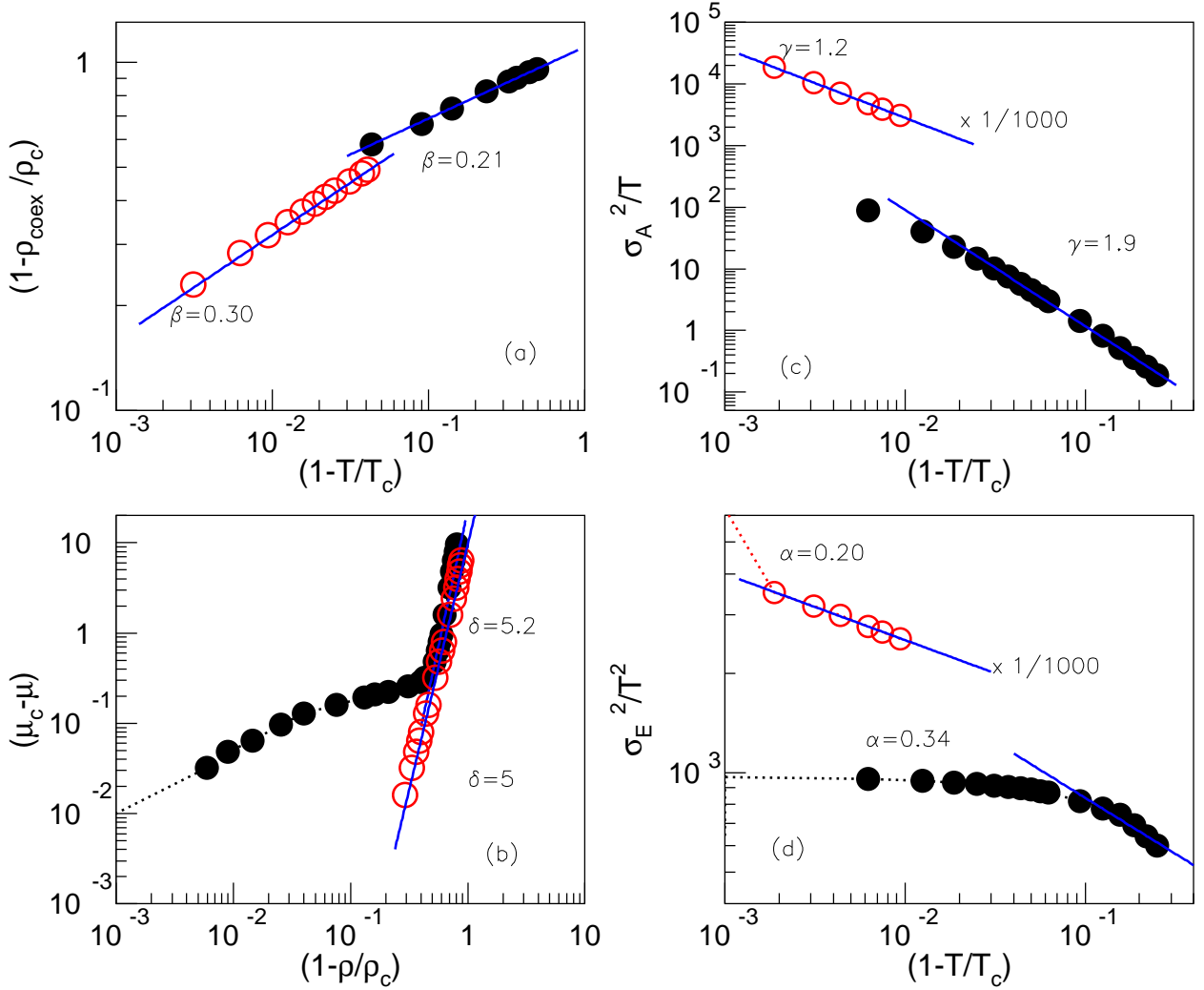}
\end{center}
\caption{(color online) 
The same as in Fig. \ref{fig:critexp_thermo_standard}
for the case in which bulk energy depends on temperature.
Results of Metropolis calculation (solid circles) correspond to $A_0$=2000.}
\label{fig:critexp_thermo_t}
\end{figure}

\begin{figure}
\begin{center}
\includegraphics[angle=0, width=0.6\columnwidth]{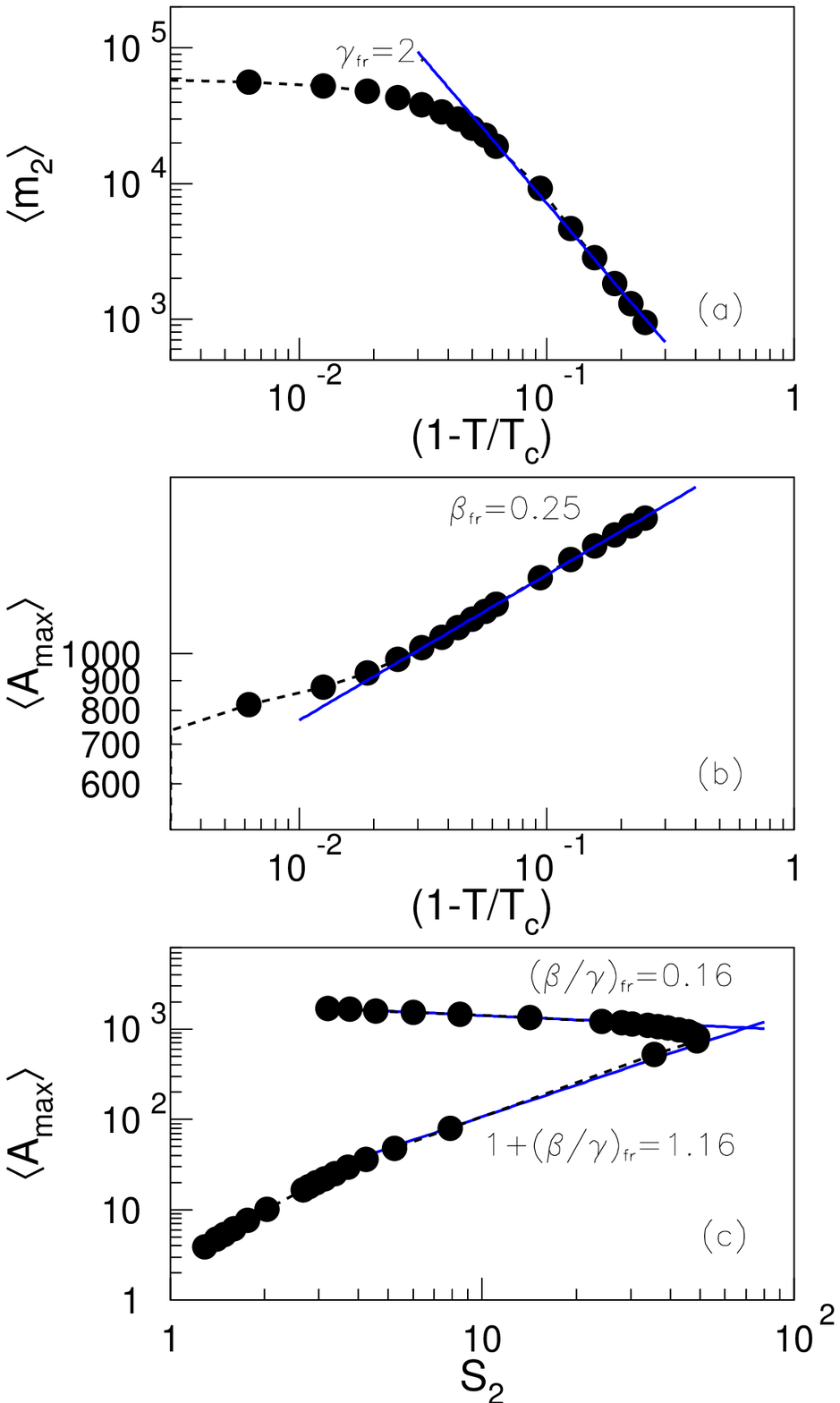}
\end{center}
\caption{(color online) 
The same as in Fig. \ref{fig:critexp_frag_standard}
for the case in which bulk energy depends on temperature.
Results of Metropolis calculation are represented with solid circles and
$A_0$=2000.}
\label{fig:critexp_frag_t}
\end{figure}

In this study we have always considered completely incompressible point-like fragments.
In this section we focus on what happens to the thermodynamics of a
one-component clusterized system if, in addition to $a_s$, also $a_v$ vanishes
while approaching $T_C$. 
From the physical point of view, no microscopic calculation suggests that this can be the case, however 
this simplified calculation may give a glimpse on what could happen if compressibility effects were 
realistically introduced, leading to a reduced bulk energy at finite temperature.
The phase diagram obtained as described in Sec. IIC is represented in Fig.
\ref{fig:phd_fisher_t} with solid circles. The same symbols are used in Figs.
\ref{fig:critexp_thermo_t} and
\ref{fig:critexp_frag_t} to depict the evolution of different thermodynamical
and fragment observables out of which critical exponents 
are calculated as described in Sec. IID.

The first result is that, when plotted in $T-\rho$, the phase diagram
superimposes perfectly on the top of the one corresponding to standard
fragment definition. This is however not valid in $T-E$ representation, 
because of the different fragment energetics and, moreover, $(\mu/T)_C$=0. 
The stability of the $T-\rho$ phase coexistence borders suggests that
the critical exponents $\beta$ and $\delta$ extracted out of the curvature
of phase coexistence and shape of the critical isotherm have the same
values as before. The confirmation is given in Fig. \ref{fig:critexp_thermo_t},
panels a and b.
Panel d shows that neither the divergence of $\ln(\sigma^2_E/T^2)$ vs. $\ln(1-T/T_C)$
is affected. By contrast, $\ln(\sigma^2_A/T)$ vs. $\ln(1-T/T_C)$ gets a linear
dependence over a broader region and $\gamma$=1.9.
Hyperscaling relations are still violated and critical behavior does not hold.
Fig. \ref{fig:critexp_frag_t} shows 
that critical exponents extracted out of fragment
properties are in reasonable agreement with the values calculated from
the thermodynamic behavior.
In this case, $f(T)= T^{-1}\left [1 -\left( 1- T/T_C \right)\Theta(T_C-T) \right ]$.

\begin{figure}
\begin{center}
\includegraphics[angle=0, width=0.9\columnwidth]{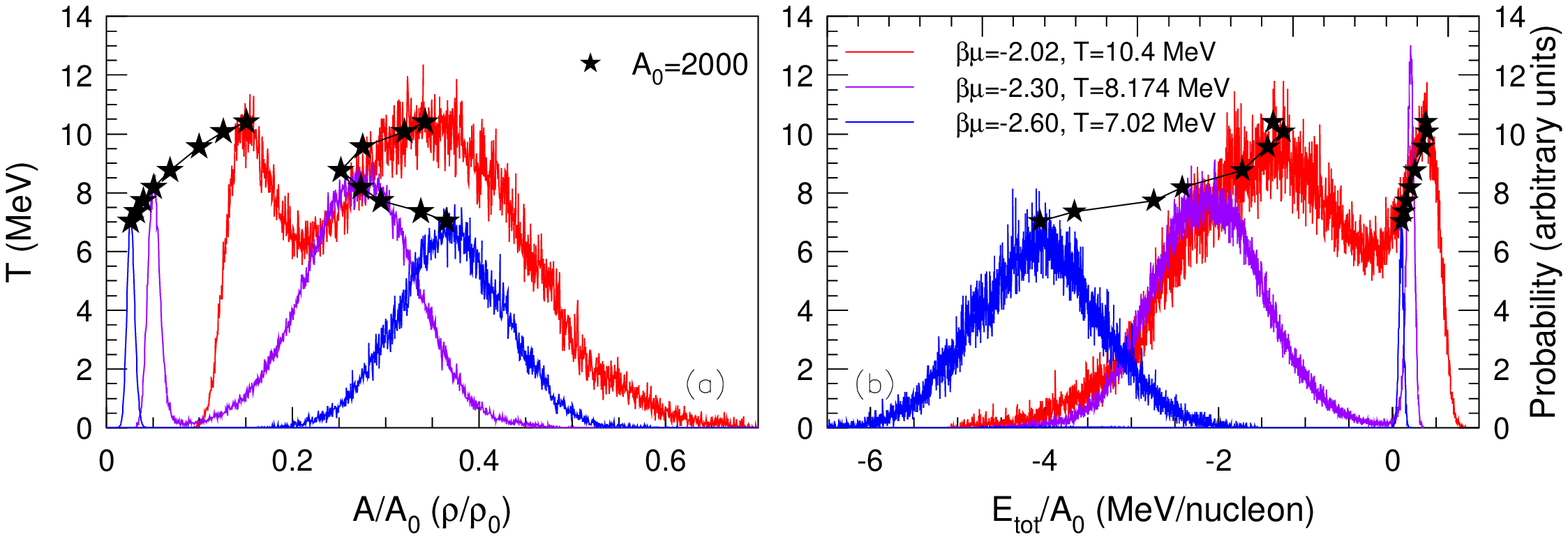}
\end{center}
\caption{
(color online)
The phase diagram in temperature - density (left panel) 
and temperature - total energy representations as obtained out
of the bimodality of corresponding distributions
for a system characterized by $V$=14476.4 fm$^3$ ($A_0$=2000)
within the grand-canonical ensemble.
A free volume correction is included according to 
Eq. (\ref{eq:exact_freevol}).}
\label{fig:phd_exactfreevol}
\end{figure}

{
\subsection{F. Free volume correction}

Working under the point-like cluster hypothesis, as we did so far,
the system's thermodynamics is entirely dictated by fragment energetics.
More sophisticated cluster-based statistical models include, in the spirit of
Van der Waals theory of fluids, the geometrical effect of 
clusters finite size which effectively act as a repulsive
fragment-fragment interaction for incompressible clusters.

To understand which is the role the excluded volume effect plays in the 
present model, let us consider what happens if each cluster forbids
for all the other clusters a volume equal with its own volume.
Mathematically, this means that the factor $V^{N_k}$ entering in the 
Eq. (\ref{eq:Wc}) should be replaced with
\begin{equation}
 V^{N_k} \rightarrow V(V-V_1)(V-V_1-V_2)...(V-V_1-V_2-...-V_{N_k-1}),
\label{eq:exact_freevol}
\end{equation}
where $V_i$ is the intrinsec volume of fragment $i$, here considered as being
spherical, $V_i=4 \pi/3 r_0^3 A_i$.

Fig. \ref{fig:phd_exactfreevol} shows the phase diagram of a $A_0=2000$
1-component system obtained as described in Sec. IIC. We note that, for the
first time, the liquid branch of the phase coexistence region is situated at
$\rho < \rho_0$ and, because of geometric effects, it bends at $T>8$ MeV.
This second behavior obviously prevents criticality studies.
}

\section{III. Application to finite systems - sensitivity to the statistical ensemble}

As we have already mentioned, Fisher cluster models have been intensively used to address
multifragmentation \cite{mekjian,dasgupta,dasgupta2006}
and the thermodynamic behavior of finite systems 
has been constantly compared to the one of their infinite counterparts.
The long-range non-saturating Coulomb interaction, which has to be included
when describing finite fragmenting nuclei, 
prevents addressing the thermodynamic limit and will not be considered here.
Interference with isospin-asymmetry effects is also avoided by 
exclusively considering isospin-symmetric systems.
Even if we ignore these effects,
realistic applications to fragmenting finite nuclear systems differ from the
model we have presented in Section II also 
because they take into account the excluded volume correction to the translational motion, do not
generally consider the topologic Fisher factor and treat the internal cluster 
level density realistically without employing the low-temperature approximation
eq.(\ref{entropy_fisher}).

These differences
suggest that the multifragmentation
phase diagram may be different from the one plotted in Fig. \ref{fig:phd_fisher_standard}.

\begin{figure}
\begin{center}
\includegraphics[angle=0, width=0.85\columnwidth]{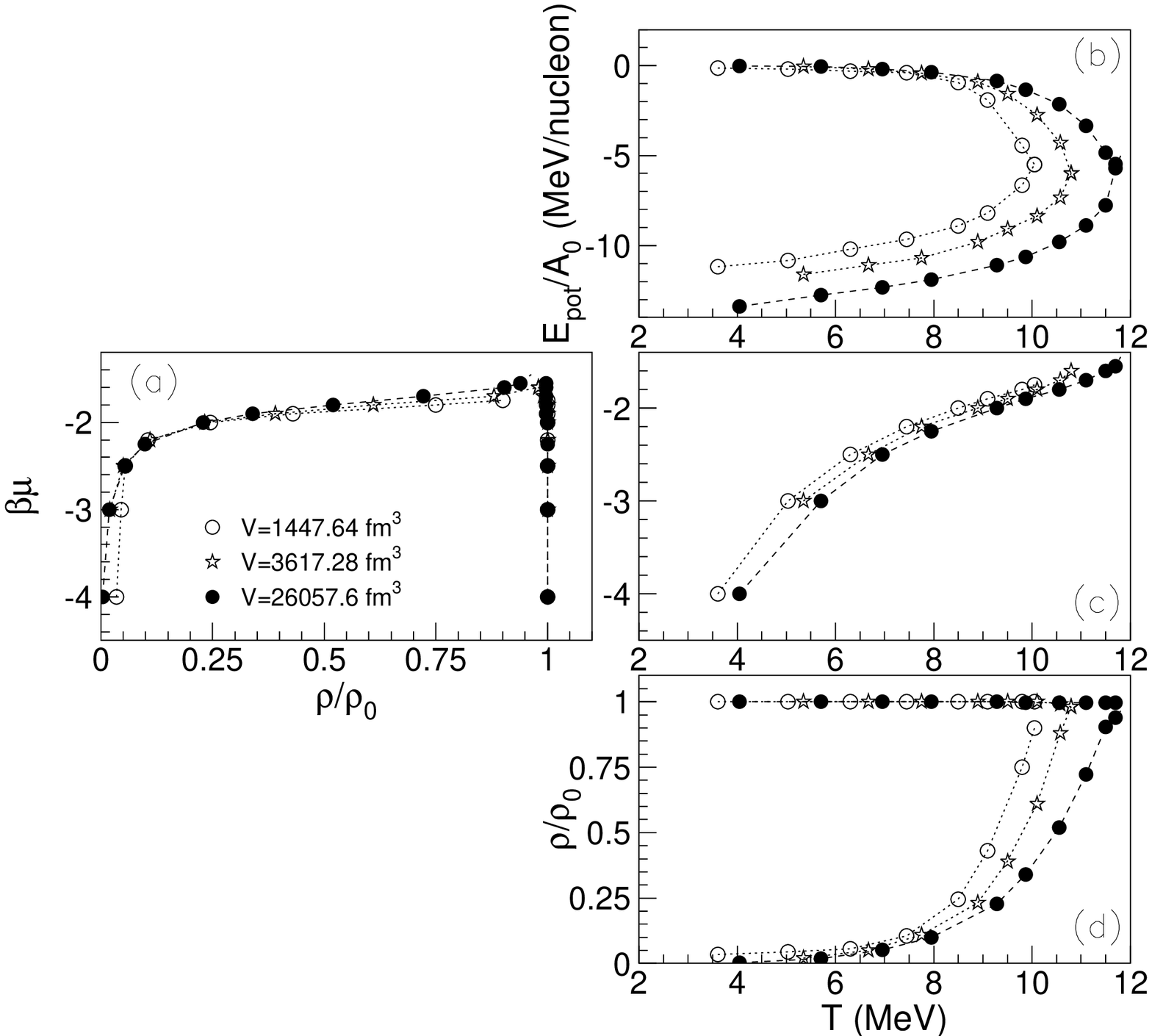}
\end{center}
\caption{
The phase diagram of the cluster model including excluded volume effects 
according to Eq. (\ref{freevolume}) as obtained out of the bimodal behavior
of the total number of particles within the grandcanonical ensemble. 
Left side: $\alpha(=\beta \mu)$ as a function of density. 
Right side: potential energy per particle, $\alpha$ 
and density as a function of temperature.
Empty circles, empty stars and solid circles correspond to charge-neutral
isospin-symmetric systems with 
volume $V$=1447.64 fm$^3$ ($A_0=200$), $V$=3617.28 fm$^3$ ($A_0$=500) and, 
respectively, $V=26057.62$ fm$^3$ ($A_0=3600$).}
\label{fig:grandcanonic}
\end{figure}

To see whether it is the case, the phase diagrams of
different charge-neutral isospin-symmetric systems ($A_0$= 200, 500, 3600)
obtained as described in Section IIC 
are depicted in Fig. \ref{fig:grandcanonic} with open circles, open stars and,
respectively, solid circles.
Standard nuclear multifragmentation fragment definition is employed
\cite{mmm}. Parameters of fragment binding energy have values close to the one used in
Section II, $a_v$=15.4941 MeV, $a_s$=17.9439 MeV,
and we have additionally taken for the symmetry energy $a_I=1.7826$
\cite{ld_mass}.
Nuclear level density is taken as,
\begin{equation}
\rho(\epsilon)=\frac{\sqrt{\pi}}{12 a^{1/4}\epsilon^{5/4}}
~ \exp(2 \sqrt{a \epsilon}) ~ \exp(-\epsilon/\tau),
\label{eq:nuclrho}
\end{equation}
with  $a=0.114 A+0.098 A^{2/3}$ MeV$^{-1}$ 
and $\tau$=9 MeV \cite{iljinov}.

As one may notice, the liquid border of the
coexistence zone is stable at $\rho=\rho_0$, but the temperature of the 
second-order transition point is situated well below $T_C$=16 MeV and,
moreover, presents a system size dependence.
The shift is essentially due to the factor $A^{3/2}$ in the partition sum coming from 
the translational motion which does not vanish even at $T_C$. 
In this case, it is clear that not only critical behavior is lost, but also
the power law shape of fragment size distributions in the vicinity
of the phase diagram limiting point. 
This result is in qualitative agreement with the conclusions of 
Ref. \cite{mekjian}, where the translational energy factor in the fragment
partition sum is responsible for the diminish of the second-order transition 
temperature from 16 MeV to 7 MeV and the lost of linear shape of fragment
size distributions in log-log scale.

The system size dependence of $T_{lim}$ is due to the
inclusion of inter-fragment interactions by means of an excluded volume 
in the spirit of the Van der Waals gas  
and persists up to $A_0 \approx 4000$.
 
Indeed, if we consider that only the space not occupied by the clusters is
available for the center-of-mass motion,
the factor $V^{N_k}$ in Eq. (\ref{eq:Wc}) has to be replaced by
Eq. (\ref{eq:exact_freevol}).
Assuming for simplicity that 
$V_1=V_2=\dots=V_{N_k}=(V/n)/N_k$, which is reasonable only for
  relatively symmetric fragment partitions or high multiplicities,
Eq. (\ref{eq:exact_freevol}) may be written as $V^{N_k}\xi$ with,
\begin{equation}
\xi=\prod_{i=1}^{N_k-1}\left ( 1-\frac{i}{nN_k}\right ),
\label{freevolume}
\end{equation}
where $n=V/V_0$, see Ref. \cite{mmm} for details.
Since,
\begin{equation}
\lim_{V\to\infty, V_0 \to \infty} \xi = 1,
\end{equation}
we can see that the excluded volume effect becomes negligible only for very large
systems. This confirms that the point-particle approximation
is correct only when dealing with the thermodynamic limit.

\begin{figure}
\begin{center}
\includegraphics[angle=0, width=0.85\columnwidth]{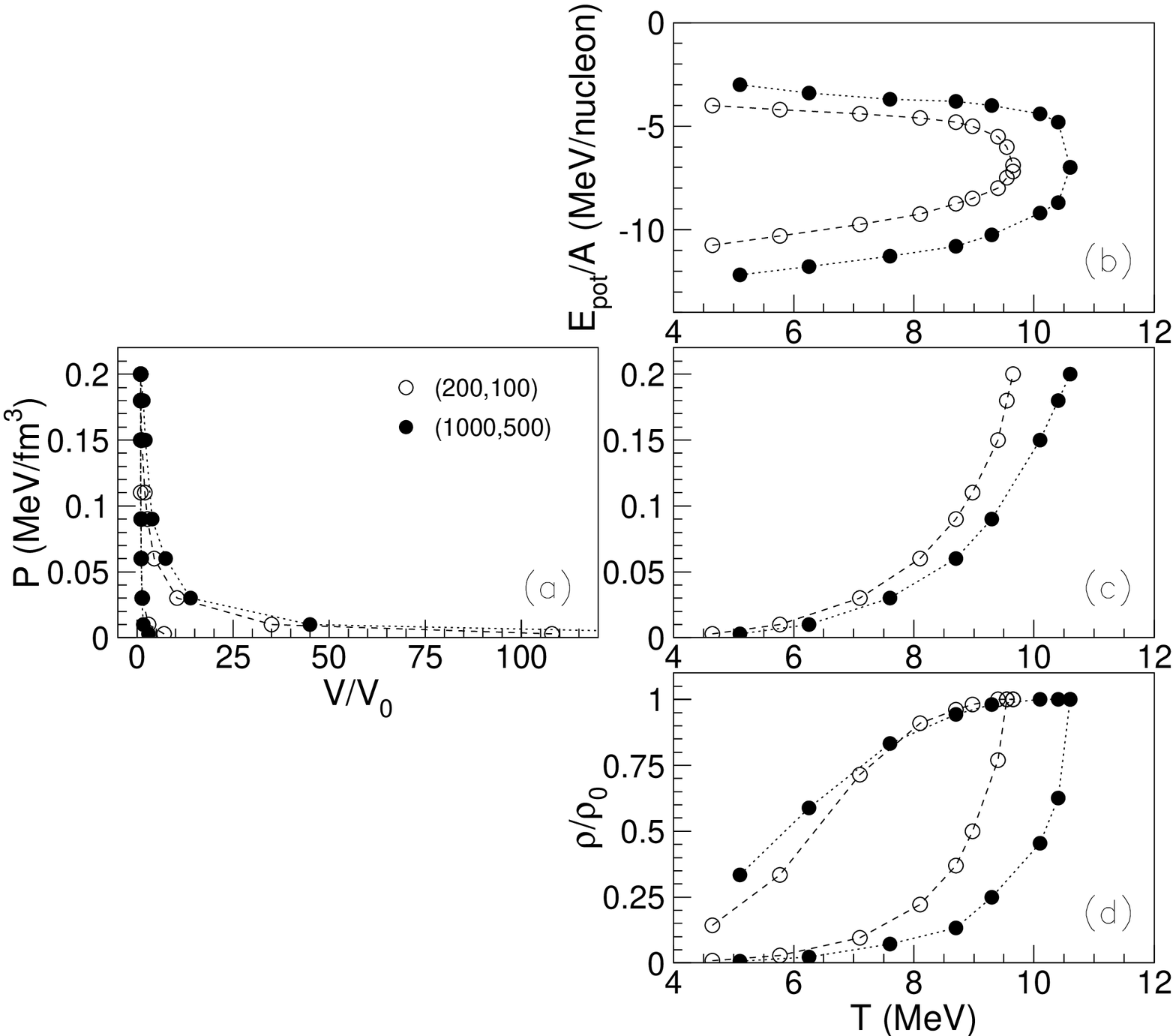}
\end{center}
\caption{
Phase diagram of the cluster model including excluded volume effects 
according to Eq. (\ref{freevolume}) in the isobar canonical ensemble.
The considered systems are neutral and isospin-symmetric. 
Left side: pressure as a function of volume. 
Right side: potential energy per particle, pressure and density as a 
function of temperature.
Empty circles correspond to a calculation with $A_0=200$,
filled circles to $A_0=1000$.
}
\label{fig:canonic}
\end{figure}

Even more important, the thermodynamics of a finite system depends on the 
statistical framework in which it is investigated.
A constant pressure canonical ensemble may be used to trace $\rho-T$
phase diagrams out of the bimodality of the volume and to immediately compare
them with the phase diagrams obtained grandcanonically out of the bimodality
of the total number of particles.
The results corresponding to two systems ($A$=200 and 1000) 
are depicted in Fig. \ref{fig:canonic} with open and solid circles
in different representations.
As above, the system size dependence of $T_{lim}$ is the consequence of
excluded volume corrections.
Contrary to the grandcanonical results, in this case the liquid border of the 
phase coexistence region extends
over the whole density domain 
and the coexistence region at low temperatures shrinks to the
vicinity of $\rho=0$. The limiting point is characterized by $\rho=\rho_0$.
As the size of the system increases, 
the constant-pressure canonical phase diagram approaches 
the grand-canonical one to which it is identical in the thermodynamic limit.
The convergence is nevertheless very slow.

\section{IV. Application to infinite systems: neutron stars}

Clusterized nuclear matter in the temperature and density domain of a liquid-gas
phase transition is predicted to be produced also in the core of supernovae and
in the inner crust of neutron stars 
\cite{ravenhall,lattimer,lattimer_swesty,watanabe2,horowitz,hartreefock,maruyama,constanca}.
In this case, the system's net charge neutrality is insured by the 
ultra-relativistic gas of electrons in which the nuclear fragments
are embedded.

In the free neutron regime 
($4.3 \times 10^{11}$ g/cm$^3 \leq \rho \leq 2.5 \times 10^{14}$ g/cm$^3$)
the matter is presently figured-out as a lattice of nuclei immersed in a
nucleon and electron gas and theoretically treated 
as a succession of non-interacting Wigner-Seitz cells 
\cite{baym71,bonche81,lattimer85}.
 
If we adopt this scenario, the thermodynamics of the whole star crust may be
investigated analyzing only one such a cell and represents a perfect
application of the cluster model.
The effect of the Coulomb energy is that of modifying the energy of cluster 
configuration, and of suppressing the critical behavior of 
nuclear matter \cite{napolitani}. 
Because of that last point, we have not included the Fisher topologic factor $-\tau \ln A$ 
in the cluster entropy. 
Concerning the modified energetics, this effect is taken into account by altering the statistical
weight of a configuration Eq. (\ref{eq:Wc}) 
according to,
\begin{eqnarray}
W_k=\frac{1}{N_k!} \exp \left ( -\beta V_C(k)\right ) V^{N_k} \prod_{i=1}^{N_k}
\left[\left(\frac{m A_i T}{2 \pi \hbar^2}\right)^{3/2} \rho(\epsilon_i) ~
w_{\beta,\mu_n,\mu_p}(A_i,Z_i) \right],
\label{eq:Wc_star}
\end{eqnarray}
where the weight associated to a given cluster $(A_i,Z_i)$ is
\begin{eqnarray}
T \ln w_{\beta,\mu_n,\mu_p}(A_i,Z_i)&=&
 \left (a_v A_i -a_s \left( 1-T f(T)\right) A_i^{2/3}\right)\left ( 1- a_I(A_i-2Z_i)^2/A_i\right ) \nonumber \\
&-&
\epsilon(A_i,Z_i)
+\mu_n (A_i-Z_i)+\mu_p Z_i,
\label{eq:wi_star}
\end{eqnarray}
and nuclear fragments are described as in Sec. III.
The Coulomb energy may be calculated for each configuration as \cite{lattimer85},
\begin{equation}
V_C(k)=\sum_{i=1}^{N_k} \frac35 c(\rho) \frac{e^2 Z_i^2}{r_0 A_i^{1/3}},
\end{equation}
with
\begin{equation}
c(\rho)=1-\frac32 \left(  \frac{\rho_e}{\rho_{0p}} \right)^{1/3}+
\frac12 \left(\frac{\rho_e}{\rho_{0p}} \right),
\end{equation}
accounting for the screening effect of electrons.
$\rho_{0p}=Z/A \rho_0$ denotes the proton density inside the nuclei and
$\rho_e$ is the electron density.

\begin{figure}
\begin{center}
\includegraphics[angle=0, width=0.8\columnwidth]{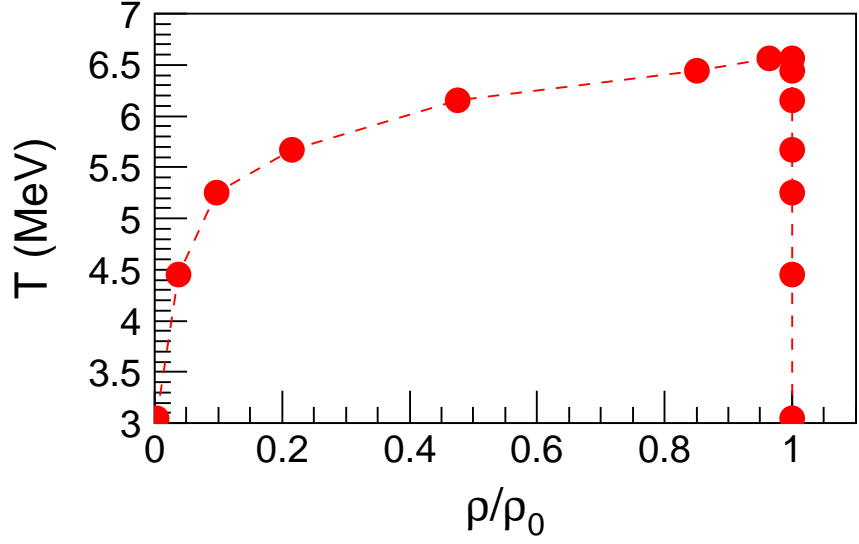}
\end{center}
\caption{(color online) 
Grand-canonical phase diagrams in temperature - density representation 
for a Wigner-Seitz cell of volume $V$=28952.91 fm$^3$.
}
\label{fig:phd_starmatter}
\end{figure}

Fig. \ref{fig:phd_starmatter} illustrates the phase diagram of a Wigner-Seitz cell
of volume $V$=28952.91 fm$^3$ ($A_0$=4000) as obtained from the bimodality of
total particle number distributions in a grandcanonical ensemble.

One may directly compare the obtained phase diagram with the one corresponding
to a system of similar size where Coulomb is switched-off 
(solid symbols in Fig. \ref{fig:grandcanonic}).
We can see that the shape of the phase diagram does not change. 
This would not necessarily be true any more 
if the weight of the translational motion $\propto (VAT)^{3/2N_k}$ at high temperature was reduced 
through a temperature dependent reduction factor as in the Lattimer-Swesty
equation of state \cite{lattimer}.  
However the prescription for such a term is completely phenomenological, and we will not employ it here.
A clear effect is seen on the limiting temperature, which  is diminished by several MeV,
in agreement with results obtained in Ref. \cite{coulomb} for small systems.
An opposite trend has been obtained within a
classical microscopic model for Coulomb frustration \cite{napolitani}, and
the understanding of this discrepancy requires further analyses. 

As observed in section II, the fact that the high density transition line lies
along $\rho=\rho_0$ is not physical
and is due to the incompressibility assumption of the dense phase which is
here represented as a single uncharged nucleus
of infinite size. This line cannot therefore be interpreted as a prediction
for the crust-core transition density as a function of temperature. 
Moreover, at densities close to saturation,
exotic pasta phases are known to appear
\cite{watanabe2,horowitz,hartreefock,maruyama,constanca} 
and the cluster energy functional should be modified to account for these non-spherical geometries.
However in the low density region this model should give a good approximation
of the thermodynamics of proto-neutron
stars and supernova matter \cite{ohnishi,lattimer_swesty,mishustin}. 
The evaluation of the equation of state with the 
inclusion of gammas and neutrinos is in progress.

\section{V. Conclusions}

In conclusion, the thermodynamics of clusterized matter was investigated in
connection with a liquid-gas phase transition in nuclear matter.
The phase diagram presents first and second order phase transitions and 
may be derived out of the grand-canonical bimodality structure of total particle number, total
energy and largest cluster in each event.
Results of an exact Metropolis simulation are systematically compared with
predictions of the analytical Fisher model which by construction is critical.
At variance with the analytical counterpart,
the predictions of the exact model indicate that the liquid phase is characterized by
$\rho=\rho_0$ and that vanishing fragment surface energy is not a sufficient condition to
obtain criticality. 

The clusterized system model is applied to  charge-neutral isospin-symmetric finite systems 
and special attention is given
to the way in which the system thermodynamics depends on the employed statistical framework. 
Clusterized systems with net charge neutrality may be used also to mimic star matter.
A first schematic calculation of the phase diagram in the Wigner-Seitz cell approximation 
is built and it is shown that within this approximation charge fluctuations lead
to the decrease of the limiting temperature. 

\section{acknowledgements}
Ad. R. R acknowledges partial support from the Romanian National Authority for Scientific
Research under PNCDI2 programme, grant {\it IDEI nr. 267/2007} and kind
hospitality from LPC-Caen within IFIN-IN2P3 agreement nr. 07-44.
F.G. acknowledges partial support from ANR under Project NExEN.

\end{document}